\documentclass[10pt, aps, prl, superscriptaddress, noshowkeys, twocolumn,
floatfix, preprintnumbers,showpacs]{revtex4-1}

\usepackage{amsmath, amssymb, amsfonts}
\usepackage{graphicx, color}
\usepackage{times}

\usepackage{graphicx}
\usepackage{bm,color}
\usepackage[normalem]{ulem}

\newcommand{\bk}{{\bf{k}}}
\newcommand{\bp}{{\bf{p}}}

\newcommand{\br}{{\bf{r}}}
\newcommand{\bK}{{\bf{K}}}

\allowdisplaybreaks
\draft % marks overfull lines with a black rule on the right

\begin{document}
\title{{ Mass Renormalization in Transition Metal Dichalcogenides  }}

\author{T.~Stroucken}%\footnote{Author to whom correspondence should be addressed: tineke.stroucken@physik.uni-marburg.de}
\author{J.~Neuhaus}
\author{S.~W.~Koch}
\affiliation{Department of Physics and Material Sciences Center, Philipps University Marburg, Renthof 5, D-35032 Marburg, Germany}

\begin{abstract}

It is shown that the three-fold rotational symmetry in transition metal dichalcogenides leads to a Coulomb induced renormalization of the effective electron and hole masses near the $K$-points of the Brillouin zone. The  magnitude of the renormalization depends on the dielectric configuration. The effective exciton mass $m=0.4 m_0$ of a freely suspended MoS$_2$ monolayer changes to $m= 0.35 m_0$ with hBN encapsulation. The mass renormalization increases the excitonic binding energy and reduces the exciton diamagnetic shift and cyclotron frequency. Detailed comparisons with high field measurements of the excitonic diamagnetic shift show excellent agreement.

\end{abstract}
\date{\today}


\maketitle 
%\maketitle must follow title, authors, abstract and \pacs


With the ability to create them as monolayers, van der Waals bonded layers have
emerged as a new material class, including graphene and transition metal
dichalcogenides (TMDCs).  In particular, TMDC monolayers have attracted
considerable attention, partly because of their extraordinary strong
light-matter interaction and excitonic effects, but also because of their potential application in valleytronic devices.  

Common to TMDCs and other layered van der Waals materials is the arrangement of
the atoms within the layers  into a honeycomb lattice, with the atoms located
at the corners of the hexagon. In reciprocal space, the first  Brillioun zone
also has honeycomb geometry with direct band gaps occurring at the six corners
of the hexagon. Since neighboring  $K$ and $K'$ valleys are related by the
parity or time reversal transformation, these are addressed with oppositely circularly polarized light.
This fascinating feature has lead to fascinating new physics, in particular the novel concept of valleytronics\cite{Xiao2012,Molina-Sanchez2013,Cao2012,Zeng2012,Mak2012,Zeng2013}.

Whereas the valley dependent optical selection rules are correctly predicted by
density functional theory (DFT), the optical selection rules and the coupled
spin-valley dynamics has been explained in a very elegant and intuitive manner
by assigning relativistic quasi-particles with a pseudo-spin to the $K$ and
$K'$ valleys. Since its proposal in the original work \cite{Xiao2012}, the so
called massive Dirac-Fermion model (MDF) Hamiltonian has been very successfully
applied to describe many of the near $K$ point electronic and optical TMDC
properties \cite{li2013,konabe2014,zhang_2014,berkelbach_2015,Yu2019,Tahir_2014,Li2020,Henriques2020,Hyn2020,Zollner2020,Shah2020,doi:10.1142/S021798492050181X,doi:10.1063/1.5118327,Bai2019,Dong2019}.
\cite{Li2020}
However, as we show in this Letter, the MDF Hamiltonian does not display the
full three-fold rotational symmetry of the lattice, nor does it properly account for the change of orbital angular momentum of the basis functions upon absorption. We therefore propose a modified Hamiltonian (mMDF Hamiltonian) that incorporates the three-fold rotational symmetry. With this mMDF Hamiltonian, we obtain a Coulomb induced renormalization of the effective electron and hole masses near the K-point that increases the excitonic binding energy and reduces the exciton diamagnetic shift. 

In its original form, the MDF Hamiltonian is given by
\begin{equation}
H_0 =\sum_{\tau s\bk}\Psi^\dagger_{\tau s\bk} \left(
\begin{array}{cc}
\frac{\Delta_{s\tau}}{2}&\tau\hbar v_F k e^{-i\tau\theta_\bk}\\
\tau\hbar v_F k e^{i\tau\theta_\bk}&-\frac{\Delta_{s\tau}}{2}
\end{array}
\right)\Psi_{\tau s\bk},
\label{Hamiltonian}
\end{equation}
where $\Delta_{s\tau}$ is the spin and valley dependent gap, $v_F=ta$ is the Fermi-velocity, $a$ is the lattice constant, $t$ is an effective hopping matrix element, and $\Psi_{\tau s\bk}$ are two-component pseudospinors spanned by the $d$-type Mo-basis functions
$|d_z^2\rangle$ und $\left(|d_{x^2-y^2}\rangle +i\tau |d_{xy}\rangle\right)/\sqrt{2}$.
Eigenstates of $H_0$ have the relativistic dispersion $\epsilon_{s\tau k}=\sqrt{\left(\frac{\Delta_{s\tau}}{2}\right)^2+(\hbar v_F k)^2}$ and can be chosen such that they are simultaneous eigenstates of the operator
$\hat j_z=\hat L_z+\frac{\tau}{2}\hat\sigma_z$:
\begin{equation}
\Psi_{s\tau  \bk}^j=
\left(
\begin{array}{c} 
%u_k 
\psi_A(k){\rm e}^{i(j-\tau/2)\theta_\bk}\\
%\tau v_k
\psi_B(k){\rm e}^{i(j+\tau/2)\theta_\bk}
\end{array}\right). 
\label{spinors}
\end{equation}
Here, $\hat L_z=-i\hbar\frac{\partial}{\partial \theta_\bk}$ and $\hat \sigma_z$ is the z-component of the pseudo-spin.
Hence, the light-matter interaction is given by $H_I=-e\frac{v_F}{c}\sum_{s\tau\bk}\Psi_{s\tau  \bk}^\dagger\left(A^{-\tau}\hat \sigma^++A^{\tau}\hat \sigma^-\right)\Psi_{s\tau  \bk}$
with $A^\tau=A_x+i\tau A_y$ and $\hat \sigma^\pm$ are the ladder operators. The optical selection rules are then given by
$\langle \Psi_j|H_I|\Psi_{j'}\rangle \propto
A^{-\tau}\delta_{j,j'+\tau}+A^{\tau}\delta_{j,j'-\tau}$, showing that excitation with right(left)-handed circular polarized light increases (decreases) the angular momentum $\hat j_z$ by $\hbar$. 

Since the MDF Hamiltonian in Eq. \ref{Hamiltonian} does not display the full three-fold lattice rotational symmetry and does not account for the angular momentum change $-2\tau\hbar$ of the basis functions upon absorption, we propose to modify Eq. \ref{Hamiltonian}
by  incorporating the three-fold rotational symmetry via
\begin{equation}
H_{\rm mMDF} =\sum_{\tau s\bk}\Psi^\dagger_{\tau s\bk} \left(
\begin{array}{cc}
\frac{\Delta_{s\tau}}{2}&\tau\hbar v_F k e^{-3i\tau\theta_\bk}\\
\tau\hbar v_F k e^{3i\tau\theta_\bk}&-\frac{\Delta_{s\tau}}{2}
\end{array}
\right)\Psi_{\tau s\bk}.
\label{Hamiltonian3}
\end{equation}
This modified massive Dirac Fermion (mMDF) Hamiltonian has the same
relativistic dispersion as the original MDF Hamiltonian and the eigenstates have the general form
\begin{equation}
\tilde\Psi_{s\tau  \bk}^j=
\left(
\begin{array}{c} 
\psi_A(k){\rm e}^{i(j-3\tau/2)\theta_\bk}\\
\psi_B(k){\rm e}^{i(j+3\tau/2)\theta_\bk}
\end{array}\right). 
\label{spinors}
\end{equation}
Using the minimal substitution, the light-matter interaction is obtained as
\begin{equation}
   \tilde H_I=-e\frac{v_F}{c}\sum_{s\tau\bk}\tilde\Psi_{s\tau
   \bk}^\dagger\left(\frac{(A^{-\tau})^3}{A^2}\hat
\sigma^++\frac{(A^{\tau})^3}{A^2}\hat \sigma^-\right)\tilde\Psi_{s\tau  \bk},
\end{equation}
yielding the optical excitation rules $\langle \tilde\Psi_j|\tilde H_I|\tilde\Psi_{j'}\rangle \propto (A^{-\tau})^3/A^2\delta_{j,j'+3\tau}+(A^{\tau})^3/A^2\delta_{j,j'-3\tau}$. Hence, excitation with $\sigma^\pm$ polarized light increases the angular momentum associated with the geometric phase by $\pm 3\hbar$, while simultaneously the angular momentum associated with the basis functions is decreased by $\pm 2\hbar$, in agreement with the conservation of the total angular momentum. Furthermore, for linearly polarized light, a rotation of the polarization angle by $2\pi/3$ does not alter the optical spectra, thus reflecting the three-fold lattice symmetry.

Although the eigenstates of our new mMDF Hamiltonian have the same unrenormalized dispersion as $H_0$, the Coulomb matrix elements contain the electron-hole overlap matrix elements and hence differ in their geometric phases. In particular, the electron-hole Coulomb matrix element relevant for the description of the excitonic properties is given by
\[
W^{c\nu\nu
c}_{\bk-\bk'}=|u_ku_{k'}+v_kv_{k'}e^{-im(\theta_\bk-\theta_{\bk'})}|^2\, W_{\bk-\bk'} .
\]
Here, $W_{\bk-\bk'}$ is the screened quasi-two dimensional Coulomb potential
and $u_k^2=(\epsilon+\frac{\Delta}{2})/2\epsilon$,
$v_k^2=(\epsilon-\frac{\Delta}{2})/2\epsilon$, and $m=1$ and $m$=3 for the MDF
and mMDF Hamiltonian respectively.
As a consequence, the Coulomb-renormalized band structure can be computed from the modified gap equations\cite{stroucken2017,meckbach2018a}
\begin{eqnarray}
\label{gap1}
\tilde{\Delta}_{\mathbf{k}} &=& \Delta+\frac{1}{2}\sum_{\mathbf{k'}}\,W_{\mathbf{|k-k'|}}\,		
	\frac{\tilde{\Delta}_{\mathbf{k'}}}{{\cal{E}}_{\mathbf{k'}}},
%\left(1-f^e_{s\tau\bk'}-f^h_{s\tau\bk'}\right),%\nonumber
\\
\tilde{v}_{\mathbf{k}} &=& v_{F} + \frac{1}{2}\sum_{\mathbf{k'}}\,W_{\mathbf{|k-k'|}}\,
	\frac{k'}{k}\frac{\tilde{v}_{\mathbf{k'}}}{{\cal{E}}_{\mathbf{k'}}} 
	\cos(m(\theta_{\mathbf{k}}-\theta_{\mathbf{k'}})),
%\left(1-f^e_{s\tau\bk'}-f^h_{s\tau\bk'}\right),
\label{gap2}
\\
{\cal {E}}_\bk&=&\sqrt{\tilde{\Delta}^2_{\mathbf{k}}
+4\hbar^2\tilde{v}_{\mathbf{k}}k^2},
\label{eps_renorm}
\end{eqnarray}
with $m=3$ instead of $m=1$.

To calculate the renormalized dispersion from the gap equations, we computed the band structure and dipole-matrix elements via density functional theory (DFT)~\cite{Kohn_PR140_1965} utilizing the \emph{Vienna ab initio simulation package} (VASP)~\cite{Kresse_PRB47_1993,
 Kresse_PRB49_1994,
%  Kresse_PRB54_1996,
 Kresse_CMS6_1996}
 using the Perdew-Burke-Ernzerhof (PBE) functional~\cite{Perdew_PRL77_1996}, and including spin-orbit interaction\cite{Steiner2016}. Fitting the DFT band structure around the $K$ points by the MDF dispersion, we obtain the MDF paramaters for the gap and Fermi-velocity. The screened Coulomb potential within different dielectric environments is determined from Poison's equation using the DFT screening parameters for the parent bulk material as described in Ref.\cite{meckbach2018a}, and the Coulomb matrix elements were calculated with the aid of the DFT wave-functions. Together with the gap equations, this provides a microscopically consistent description of the renormalized quasi-particle dispersion.
\begin{figure}[h!]
\includegraphics[width =0.80\columnwidth]{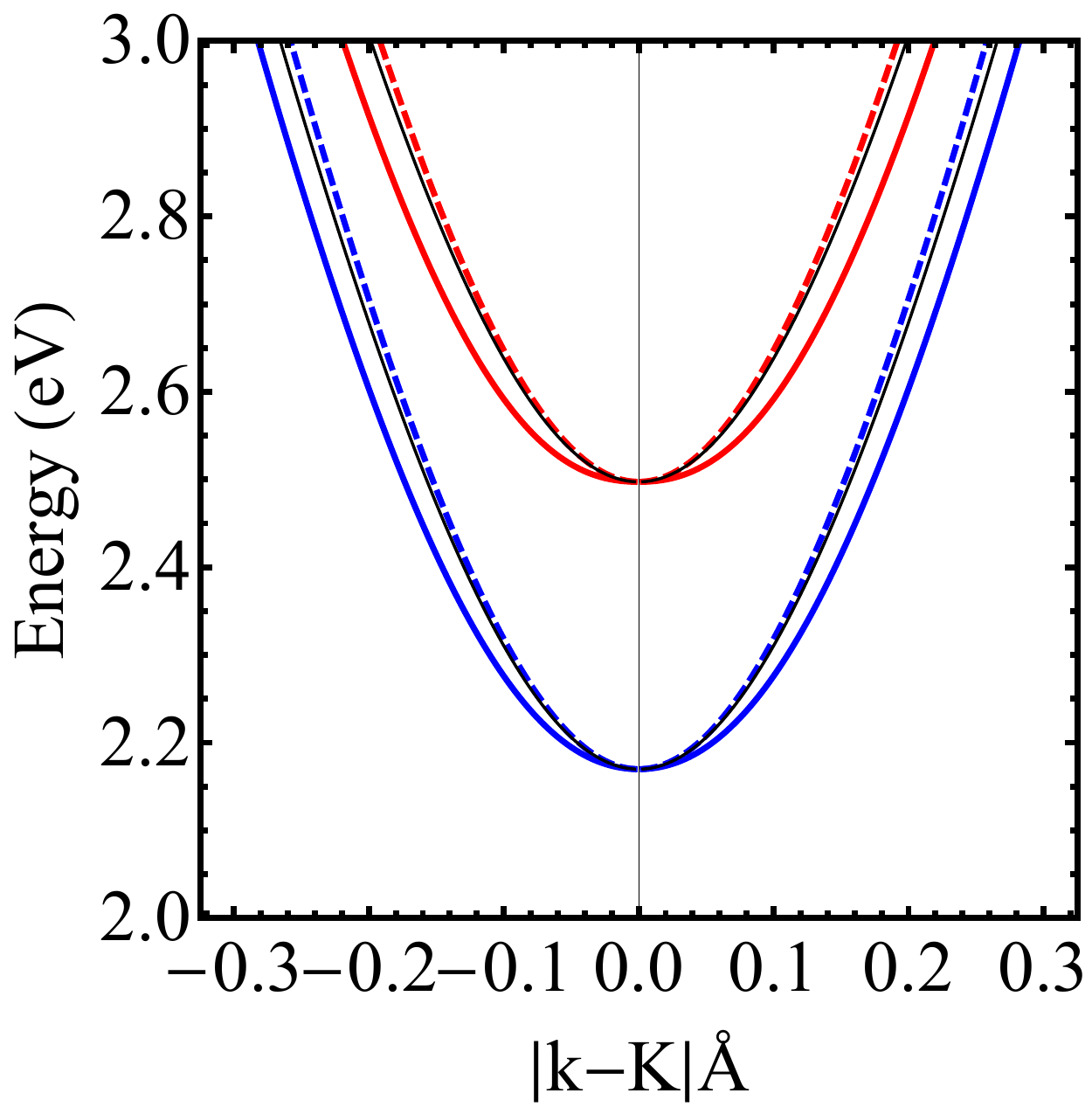}
\caption{\label{fig:7} 
%Left: 
Comparison of the renormalized dispersion at the $K$-valleys solving the gap equations with the mMDF Hamiltonian (solid lines) and the original MDF Hamiltonian (dashed lines) for the example of a suspended (red) and hBN encapsulated (blue) MoS$_2$ monolayer. The thin black lines show the unrenormalized dispersion shifted to match with the renormalized gap.
 \label{fig:MoS2Dispersion}}
\end{figure}

As shown in Ref.\cite{meckbach2018a}, within numerical accuracy the
simultaneous solution of Eqs. \ref{gap1},  \ref{gap2}, and \ref{eps_renorm}
with $m=1$ leads to rigid band shifts without dispersion modifications.
Interestingly, this aspect changes significantly once the renormalization of
the Fermi-velocity properly includes the three-fold rotational symmetry. As an
example, we compare in Fig. \ref{fig:MoS2Dispersion} the renormalized energy dispersion for a suspended monolayer of MoS$_2$ using the original MDF Hamiltonian (red dashed line) with that of the mMDF Hamiltonian (red solid line). The respective blue solid and dashed lines show the corresponding results for an hBN encapsulated monoloayer.    

\begin{figure}[hb]%[t!]
\includegraphics[width =0.80\columnwidth]{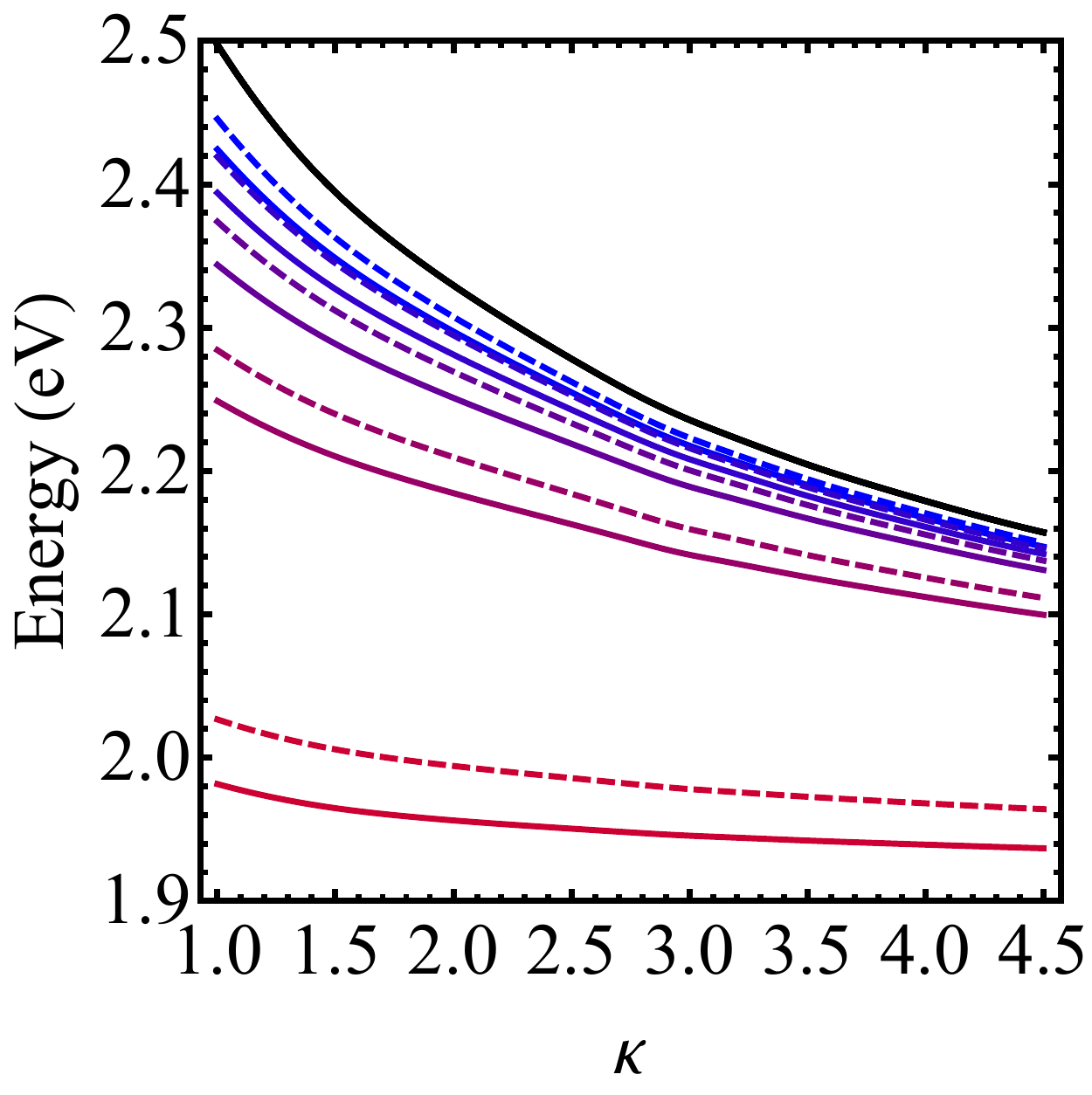}
\caption{\label{fig:7} 
%Left: Comparison of the renormalized dispersion at the $K$-valleys solving the gap equations with the mMDF Hamiltonian (solid lines) and the original MDF %Hamiltonian (dashed lines) for the example of a suspended (red) and hBN encapsulated (blue) MoS$_2$ monolayer. The thin black lines show the unrenormalized %dispersion shifted to match with the renormalized gap. Right: 
Exciton resonance energies of MoS$_2$ as function of the dielectric environment
represented by the effective dielectric constant $\kappa=(\epsilon_{\rm
top}+\epsilon_{\rm bottom})/2$. The lowest five resonances (colored) and the band gap
(black) are
shown resulting from the modified mMDF
Hamiltonian (solid lines) and the original 
MDF Hamiltonian (dashed lines).% Solid lines illustrate calculations employing
%the modi\-fied Hamiltonian ($m = 3$), whereas the dashed lines result from using
%$m = 1$.
 \label{MoS2Dispersion}}
\end{figure}

One clearly recognizes that the curvature of the mMDF dispersion at the bottom of the valleys is smaller as compared to the input DFT dispersion, i.e. the mMDF theory yields an enhanced effective mass, whereas the original MDF calculations merely yield a rigid shift of the unrenormalized dispersion.
A quadratic fit in the region $|\bk-\bK|\le 0.1\textrm{\AA}^{-1}$ yields the
effective reduced masses of $m_r^{\rm vac}=0.4\,m_0$ for the suspended monolayer
in vacuum, and $m_r^{\rm hBN}=0.35\,m_0$ for an hBN encapsulated MoS$_2$
monolayer ($m_r^{\rm vac}=0.326\,m_0$ and $m_r^{\rm hBN}=0.312\,m_0$ if fitted
on $|\bk-\bK|\le 0.2\textrm{\AA}^{-1}$).
These values are in excellent agreement with the reported effective electron
mass of $m_e\approx 2 m_r=0.7\,m_0$ extracted from Shubnikov–de Haas (SdH)
oscillations in hBN encapsulated MoS$_2$ monolayers\cite{pisoni2018} and should
be compared to the DFT value of the effective reduced mass $m_r^{\rm DFT}=0.261
\,m_0$. In the corresponding calculations for MoSe$_2$ (not shown here), we find the effective reduced masses $m_r^{\rm vac}=0.44 \,m_0$ and $m_r^{\rm hBN}=0.382 \,m_0$ for vacuum and hBN encapsulated, respectively, which is also in excellent agreement with
the effective electron mass of $m_e=0.8\,m_0$ found for hBN encapsulated MoSe$_2$ extracted from experiments\cite{larentis2018}.

In Fig. \ref{MoS2Dispersion}, we plot the energies of the lowest exciton resonances computed with the mMDF Hamiltonian as function of the dielectric environment represented by the effective dielectric constant $\kappa=(\epsilon_{\rm top}+\epsilon_{\rm bottom})/2$. 
The enhanced effective mass increases the exciton binding energy on a suspended monolayer from $471$ to $560$ meV, resulting in a resonance energy of $1.98$ eV in vacuum. For the frequently used 
quartz substrate and hBN encapsulation, we use $\epsilon_{\rm top}=3.9$ and
$\epsilon_{\rm top}=\epsilon_{\rm bottom}=4.2$ and find the $1s$-exciton resonances at $1.951$ and $1.937$ eV respectively.

To further investigate the consequences of the mass renormalizations predicted
by the mMDF model, we compute the diamagnetic shift of the exciton resonances in TMDC systems. As has been suggested by Goryca et al.\cite{goryca2019}, the slope of the diamagnetic shift provides access to the effective exciton mass and, hence, can be used to test our theoretical predictions.

In a constant, perpendicular magnetic field, the exciton equation is given by\cite{stier}
\begin{widetext}
\label{wannier}
\begin{equation}
\left(
{\cal{E}}^e_{\bk}[B]+{\cal{E}}^h_{\bk}[B]
+
\frac{eB}{2m_r c}\hat l_z+\frac{e^2B^2}{8m_r c^2}
\hat {\bf r}^2
\right)\psi_{\mu}(\bk)
-\sum_{\bk'}W^{c\nu\nu c}_{|\bk-\bk'|}\psi_{\mu}(\bk') 
=E_{\mu}\psi_{\mu}(\bk).
\end{equation}
\end{widetext}
Here, $m_r$ is the (unrenormalized) reduced mass of the electron-hole pair, $\hat l_z$ is the angular momentum operator, $\hat \br=i\nabla_\bk$ is the position operator, $W^{c\nu\nu c}_{\bk-\bk'}$ is the statically screened electron-hole Coulomb matrix element and  ${\cal{E}}^{e/h}_{\bk}[B]$ the renormalized single-particle dispersion that contains a Zeeman shift of the atomic orbitals contributing to the valence and conduction band. In general, the term $\propto\hat l_z$ leads to a Zeeman shift of the exciton states, however, for the $s$-type bright states this term does not contribute. 
The orbital angular momentum of  $d$-type orbitals with $m_z=\pm 2$ leads to a Zeeman shift of $\pm2\mu_BB$ of the valence band maxima, where $\mu_B=e\hbar/2m_0c$ is the Bohr magneton, and a corresponding splitting between the $K^\pm$ valleys. The Zeeman shift of the atomic orbitals enters the unrenormalized gap in Eq. \ref{gap1} and is slightly enhanced by the gap-renormalization, leading to g-factors with an absolute value slightly larger than $4$. 

In the low magnetic field regime, the term $\propto B^2$ can be treated perturbatively, leading  to a quadratic shift of the exciton resonance energy, which is the diamagnetic shift. In the high field limit and for a quadratic dispersion $\bp^2/2m^*_r$, the term quadratic in $B$ leads to the formation of Landau-levels with $E_n=\hbar\omega^*_c(n+1/2)$, with a cyclotron frequency 
$\hbar\omega^*_c=\sqrt{m_r/m_r^*}\,\hbar\omega_c=e\hbar B/2\sqrt{m_r m_r^*}c$.
Hence, in the high field limit the slope of the eigenvalues gives direct access
to the mass renormalization, provided the dispersion is quadratic.

\begin{figure}[h!]%[t!]
\includegraphics[width =0.40\textwidth]{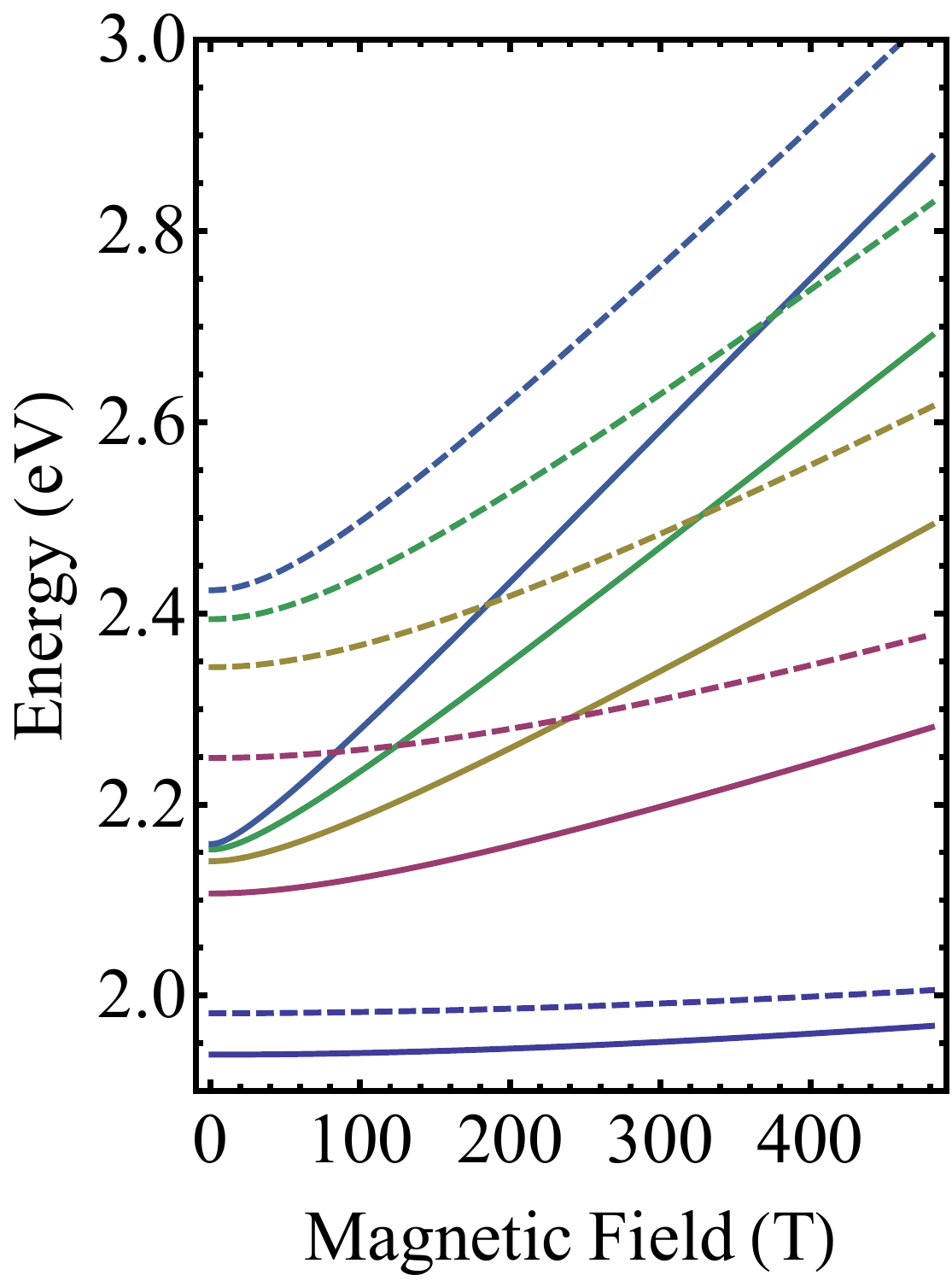}
\caption{\label{MoS2Diahighfield}  Diamagnetic shift of the five lowest A-exciton states  for a hBN encapsulated ML MoS$_2$ (solid lines) and a suspended MoS2 ML (dashed lines) for magnetic fields up to 400 T. The effective exciton mass can be estimated from the slope of the diamagnetic shift at high fields.
 }
\end{figure}
\begin{figure}[ht]%[t!]
\includegraphics[width =0.40\textwidth]{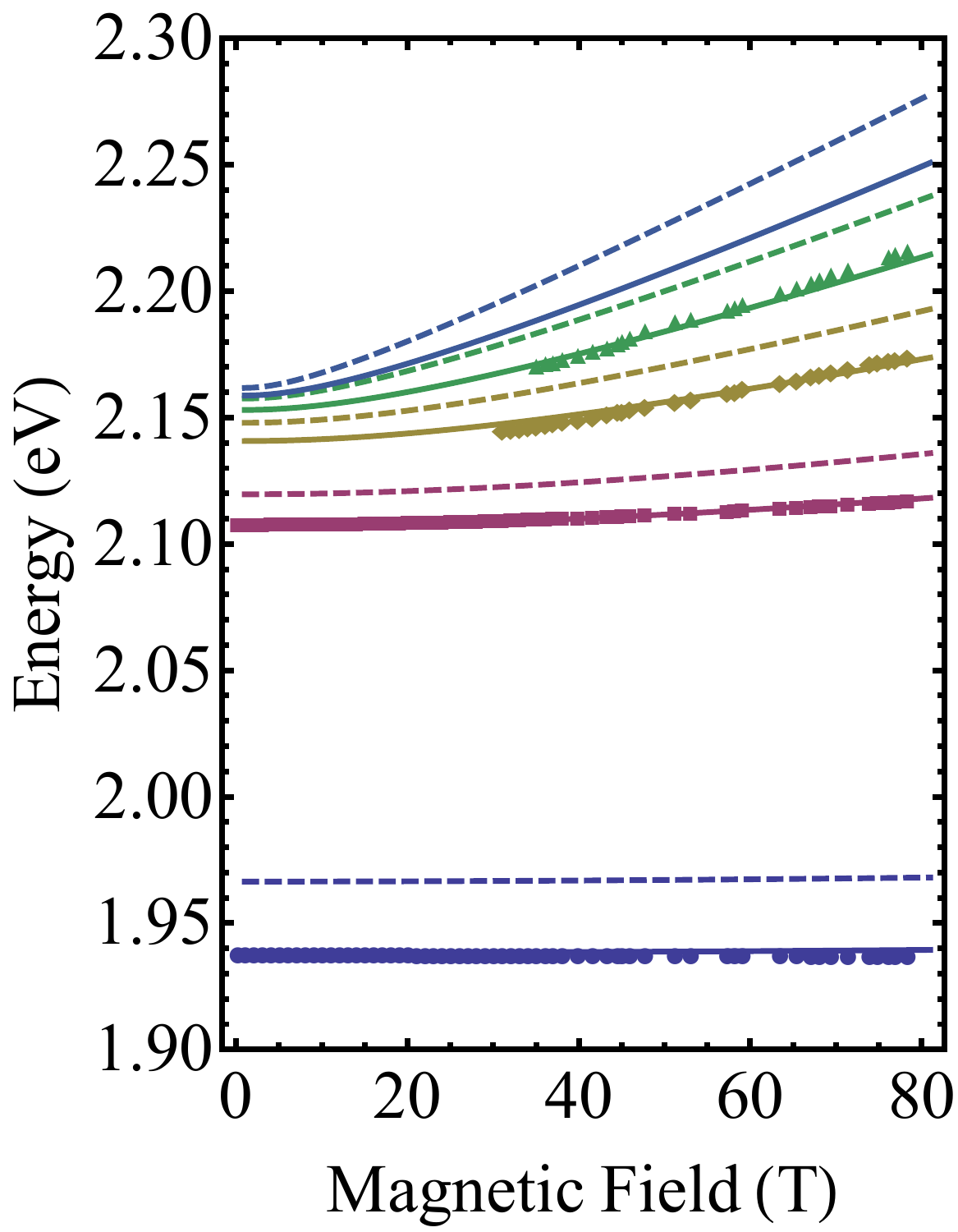}
\caption{\label{MoS2DiaII} Diamagnetic shift of the five lowest A-exciton states  for a hBN encapsulated ML MoS$_2$ with threefold rotational symmetry (solid lines) and without threefold rotational symmetry (dashed lines). The enhanced exciton mass resulting from the three-fold rotational symmetry increases the exciton binding and reduces the slope at elevated magnetic fields.
The discrete symbols denote experimental data points taken from reference \onlinecite{goryca2019}. 
 }
\end{figure}
In Fig. \ref{MoS2Diahighfield}, we compare the calculated diamagnetic shift
$\left(E_{ns}[B]+E_{ns}[-B]\right)/2$ using the mMDF dispersion for a freely
suspended MoS$_2$ monolayer (dashed lines) with that of an hBN encapsulated
configuration. The effective masses can be estimated from the slope at very
high magnetic fields. A linear fit in the region between $400$ and $500$ T
yields a renormalized exciton mass $m_r^{\rm hBN}=0.417\, m_0$ for the hBN
encapsulated sample, and $m_r^{\rm vac}=0.42\, m_0$ for the suspended monolayer, respectively. These values are in reasonable agreement with those estimated from the dispersion.

To test the reliability of our calculations, we present in Fig. \ref{MoS2DiaII} a comparison of the mMDF calculated diamagnetic shift for the hBN encapsulated MoS$_2$ monolayer (solid lines) with experimental data taken from Ref. \cite{goryca2019}. We obtain excellent agreement between theory and experiment which would not have been possible using the original MDF Hamiltonian (dashed lines). Equally good theory/experiment agreement has also been obtained for other TMDC systems, including MoTe$_2$,  WS$_2$ and WSe$_2$. Data on these materials can be found in the supporting online material.

In conclusion, we introduced a modification of the original massive Dirac
Fermion Hamiltonian that accounts for the three-fold rotational symmetry of
the TMDC lattice. We show that this modified Hamiltonian leads to a curvature
change in the computed energy band dispersion that depends on the dielectric
environment. This renormalization of the effective mass manifests itself in the
values of the predicted exciton binding energies and their magnetic field
induced shift. We demonstrate excellent quantitative agreement with
experimental data. Hence, we expect
that the proposed modified massive Dirac Fermi Hamiltonian will find widespread
use in future experimental analysis and TMDC design applications.\\
\vspace{0.9cm}
\section{Acknowledgements}

This work was funded by the DFG via the Collaborative Research Center SFB 1083. We thank Scott Croocker for giving us access to their experimental data before publication.

%https://iopscience.iop.org/article/10.1088/0953-8984/27/10/105401/pdf\end{document}
% Create reference section:

%\onecolumngrid
%%%\pagebreak
%\vspace{1cm}
%%%\section*{References}
%\twocolumngrid
%\section*{References}
%\bibliography{references_form}
%\end{document}


\end{document}